\documentclass[twoside,leqno,twocolumn,review]{article}

\usepackage[letterpaper]{geometry}

\usepackage{ltexpprt}

\author{Felix Hausberger, Marcelo Fonseca Faraj, Christian Schulz}
\title{Scalable Multilevel and Memetic Signed~Graph~Clustering} %

\newcommand{ \paragraphspacing} {0.0cm}

\usepackage[noend]{algcompatible}                           

\algnewcommand\algorithmicto{\textbf{to}}                   
\algnewcommand\RETURN{\State \textbf{return} }

\usepackage{hyperref}

\usepackage{algorithm}

\newcommand{\NEW}[1]{{\color{black} #1}}

\usepackage{multirow}
\usepackage{pdflscape}
\usepackage{numprint}
\npdecimalsign{.}

\usepackage{amsmath}
\usepackage{amsfonts}

\DeclareRobustCommand{\frcshape}{\fontfamily{frc}\selectfont}
\DeclareTextFontCommand{\textfrc}{\frcshape}

\date{}

\def\MdR{\ensuremath{\mathbb{R}}}
\def\MdN{\ensuremath{\mathbb{N}}}
\newcommand{\set}[1]{\left\{ #1\right\}}
\newcommand{\sodass}{\,:\,}
\newcommand{\setGilt}[2]{\left\{ #1\sodass #2\right\}}

\newcommand{\Id}[1]{\texttt{\detokenize{#1}}}

\newcommand{\Is}       {:=}

\newcommand{\etal}{et~al.\xspace}

\newif\ifFull
\Fullfalse

\usepackage[square,numbers]{natbib}
\usepackage[usenames,dvipsnames]{xcolor}
\usepackage{hyperref}
\usepackage{xspace}
\usepackage{todonotes}
\usepackage{booktabs}

\usepackage{afterpage}

\begin{document}

\maketitle

\begin{abstract}
In this study, we address the complex issue of graph clustering in signed graphs, which are characterized by positive and negative weighted edges representing attraction and repulsion among nodes, respectively. The primary objective is to efficiently partition the graph into clusters, ensuring that nodes within a cluster are closely linked by positive edges while minimizing negative edge connections between them. To tackle this challenge, we first develop a scalable multilevel algorithm based on label propagation and local search. Then we develop a memetic algorithm that incorporates a multilevel strategy. This approach meticulously combines elements of evolutionary algorithms with local refinement techniques, aiming to explore the search space more effectively than repeated executions. Our experimental analysis reveals that our new algorithms significantly outperforms existing state-of-the-art algorithms. 
\end{abstract}

\section{Introduction}
\label{sec:introduction}

Graphs are a powerful mathematical abstraction used to represent a wide variety of phenomena such as social networks, email exchanges, purchases, and so forth.
In many real-world applications, interactions between two entities can be accurately represented by \emph{signed graphs}, i.e., graphs containing edges with positive and negative weights.
In particular, the sign associated with an edge can indicate when the nature of an interaction between nodes is positive (e.g., attraction, similarity, friendship) or negative (e.g., repulsion, difference, animosity).
A relevant problem for the analysis of complex graphs is to find a \emph{graph clustering}, i.e., a partition of the nodes of a graph into well-characterized blocks or \emph{clusters}.
The general purpose when clustering unsigned graphs (graphs containing only positive edges) is to ensure that nodes contained in the same cluster are densely connected while nodes from distinct clusters are sparsely connected.
For clustering signed graphs, the general purpose is a bit different.
On the one hand, it is to ensure that nodes contained in the same cluster are densely connected by positive edges and sparsely connected by negative edges.
On the other hand, it is to ensure that nodes from distinct clusters are densely connected by negative edges and sparsely connected by positive edges.
A graph clustering provides information about the implicit inherent structure of a graph, which has practical applications in areas such as criminology, public health, politics, \hbox{and analysis of social networks~\cite{application_review}}.

Unsigned graph clustering has been extensively studied under various perspectives~\cite{DBLP:books/sp/social2011/Aggarwal11}. 
Signed graph clustering differs from unsigned graph clustering, requiring specific metrics and approaches.
On the one hand, traditional metrics like modularity and conductance cannot address negative edges. 
On the other hand, negative edges can make the clustering structure more explicit.
As a consequence, metrics that are not adequate to evaluate the quality of an unsigned clustering can be sufficient to evaluate a signed clustering.
One of such metrics is \emph{edge-cut}, which measures the sum of edge weights between clusters.
Unlike in unsigned graphs, where the minimum edge-cut value of zero is achieved by grouping all nodes together, the edge-cut in signed graphs is bounded by the sum of all negative edges. 
A clustering with this edge-cut value separates positive edges inside clusters and cuts negative edges, fulfilling the \hbox{aim of signed graph clustering}.

The problem of finding a clustering of signed graphs with minimum edge-cut is NP-hard~\cite{DBLP:phd/dnb/Wakabayashi86}, hence heuristic algorithms are used in practice.
The signed graph clustering problem has been solved for distinct objective functions using a wide variety of methods, including spectral~\cite{kunegis2010spectral,gallier2016spectral,knyazev2018spectral,mercado2019spectral}, evolutionary~\cite{memeticsiggraphcommdetec2020,zhu2018novel,abdulrahman2020enhanced}, combinatorial~\cite{bailoni2019generalized,HuaRandomWalk2020}, and exact integer linear~\cite{grotschel1989cutting,dinh2015toward,miyauchi2015redundant,miyauchi2018exact,koshimura2022concise} approaches.
Although the edge-cut objective function has already been deeply explored for some problems such as the graph partitioning problem~\cite{SPPGPOverviewPaper,MoreRecentAdvancesGP,DBLP:reference/bdt/0003S19,walshaw2000mpm,karypis1998fast,Walshaw07,kaffpa,sanders2012distributed}, very little work has been done to heuristically address it \hbox{in the context of signed graph clustering}.

\vspace*{\paragraphspacing}
\subparagraph*{Contribution.}
First, we introduce a scalable multilevel algorithm for the problem. We then thoroughly engineer a memetic algorithm specifically designed for this problem which uses this multilevel strategy to provide natural recombination operations.
We also parallelize our approach using a scalable coarse-grained island-based strategy. 
Experimental results demonstrate that our multilevel algorithm computes solutions comparable to solutions computed by the greedy agglomeration based \texttt{GAEC} algorithm while being 46\% faster on average.  Our memetic algorithm overall computes the lowest edge-cuts. It can reach the same solution quality as the previous state-of-the-art algorithm, \texttt{GAEC+KLj} (GAEC + Kernighan-Lin-based local search), on average 61 times faster. The difference in execution time increases with instance size. Our memetic algorithm can compute solutions of comparable quality up to four orders of magnitude~faster.

\section{Preliminaries}
\label{sec:preliminaries}

\vspace*{-0.15cm}
\textbf{Basic Concepts.}
\label{subsec:basic_concepts}
Let $G=(V=\{0,\ldots, n-1\},E)$ be an \emph{undirected signed graph} with no multiple or self edges allowed, such that $n = |V|$ and $m = |E|$.
Let $c: V \to \MdR_{> 0}$ be a node-weight function, and let $\omega: E \to \MdR \setminus \{0\}$ be a \emph{signed} edge-weight function.
We generalize $c$ and $\omega$ functions to sets, such that $c(V') = \sum_{v\in V'}c(v)$ and $\omega(E') = \sum_{e\in E'}\omega(e)$.
Let $E^-$ denote the edges with negative weight and $E^+$ denote the edges with positive weight, such that $E^- \cup E^+ = E$, $E^- \cap E^+ = \emptyset$, $\NEW{m}^- = |E^-|$, and $\NEW{m}^+ = |E^+|$.
Let $N(v) = \setGilt{u}{\set{v,u}\in E}$ denote the neighbors of $v$.
Let $N^-(v)$ denote the neighbors of $v$ \NEW{which} are connected to $v$ by an edge with negative weight.
Let $N^+(v)$ denote the neighbors of $v$  \NEW{which} are connected to $v$ by an edge with positive weight.
A graph $S=(V', E')$ is said to be a \emph{subgraph} of $G=(V, E)$ if $V' \subseteq V$ and $E' \subseteq E \cap (V' \times V')$. 
When $E' = E \cap (V' \times V')$, \hbox{$S$ is the subgraph \emph{induced} in $G$ by $V'$}.

Let a \emph{clustering} be any partition of $V$, i.e., a set of \emph{blocks} or \emph{clusters} $V_1$,\ldots,$V_t \subset V$ such that $V_1\cup\cdots\cup V_t=V$ and $V_i\cap V_j=\emptyset$, where $t\in[1,n]$.
An abstract view of the clustering is a \emph{quotient graph} $\mathcal{Q}$, in which nodes represent clusters and edges are induced by the connectivity between clusters.
We call \emph{neighboring clusters} a pair of clusters that is connected by an edge in the quotient graph.
A vertex $v \in V_i$ that has a neighbor $w \in V_j, i\neq j$, is a \emph{boundary vertex}.
The \emph{edge-cut} of a clustering consists of the total weight of the edges crossing clusters (also called \emph{cut edges}), i.e., $\sum_{i<j}\omega(E_{ij})$, where $E_{ij}\Is\setGilt{\set{u,v}\in E}{u\in V_i,v\in V_j}$.
Note that the sum of the negative edge weights is an absolute lower bound for the edge-cut of any clustering.
Let the \emph{signed graph clustering} (SGC) consist of obtaining a clustering of an undirected signed graph $G$ \hbox{in order to minimize the edge-cut}.

\begin{figure}[t]
	\centering
		\centering
		\includegraphics[width=0.45\textwidth]{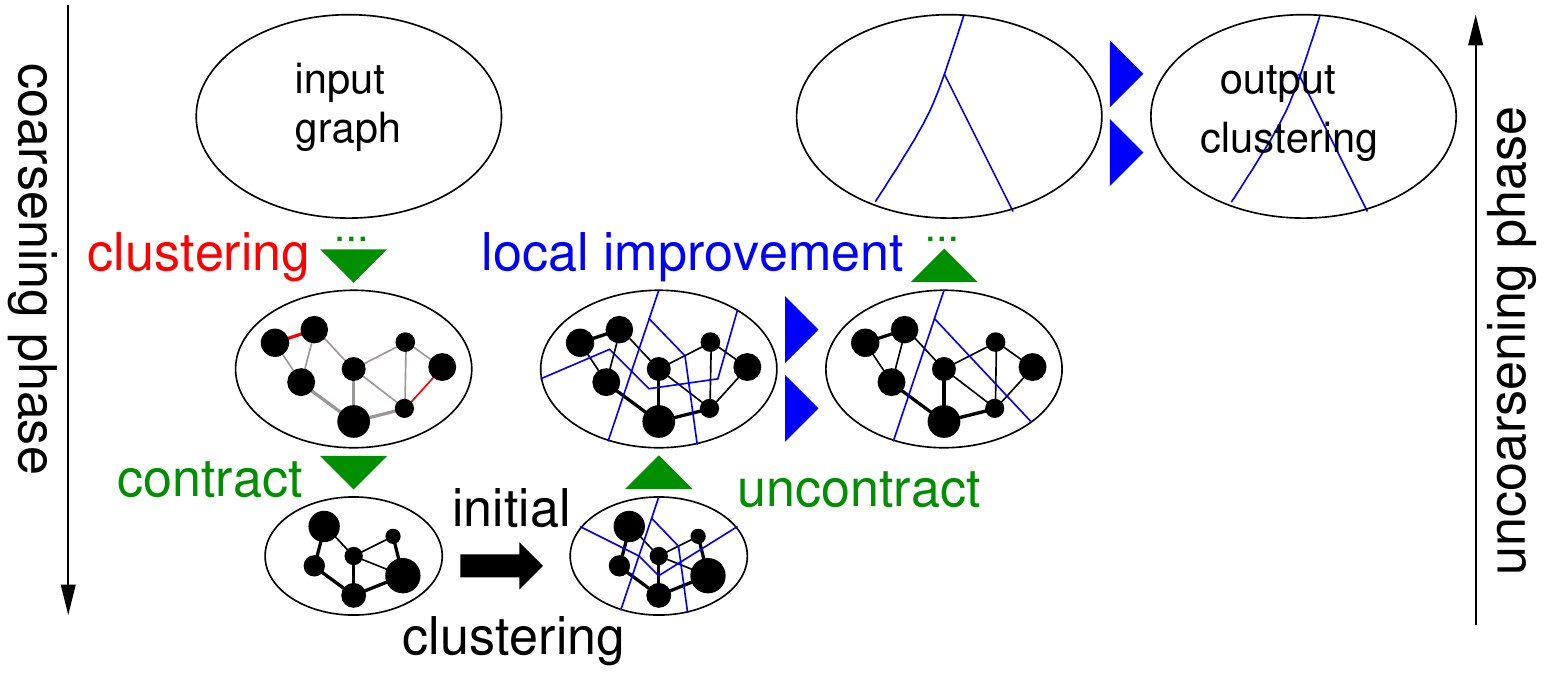}
		\caption{Multilevel scheme.}
                \vspace*{-0.25cm}
		\label{fig:MSGC}
\end{figure}

A successful heuristic for clustering or partitioning large graphs is the \emph{multilevel}~\cite{MoreRecentAdvancesGP} approach, illustrated in Figure~\ref{fig:MSGC}.
It recursively computes a clustering and \emph{contracts} it in order to \emph{coarsen} the graph into smaller graphs.
An \emph{initial clustering} algorithm is applied to the smallest (\emph{coarsest}) graph, then the contraction is undone.
At each level, a \emph{local search} method is used to improve the partitioning induced by the coarser level. 
\emph{Contracting} a cluster of nodes $C=\set{u_1, \ldots, u_{\ell}}$ involves replacing them with a new node~$v$ whose weight is the sum of the weights of the clustered nodes and is connected to all elements  $w \in \bigcup_{i=1}^{\ell} N(u_i)\setminus C$, \hbox{$\omega(\set{v,w})=\sum_{i=1}^{\ell}\omega(\set{u_i,w})$}.
This ensures the transfer of partition from a coarser to a finer level maintains the edge-cut. %
The \emph{uncontraction} of a node undoes the contraction. 
Local search moves nodes between blocks to reduce the objective. %

\emph{Evolutionary} or \emph{Memetic algorithms}~\cite{evolutionary_book} are population-based heuristics that mimic natural evolution to optimize a problem.
They use a convenient \emph{notation} to represent a setting of all decision variables, called \emph{individuals}, and evaluate their quality through a \emph{fitness} function (simulation or mathematical~\cite{sastry2014genetic}). 
A \emph{population} of individuals evolves during the algorithm. 
\emph{Recombination}~\cite{goldberg1989messy} (crossover) \emph{exploits} characteristics of previous individuals to create new and better solutions. 
\emph{Mutation}~\cite{Michalewicz:1996:GAD:229930} introduces random variations to \emph{explore} the search space and escape local optima.

\subsection*{Related Work.}
\label{subsec:related_work}

There is a huge body of research on signed graph clustering.
Here, we focus on the results specifically related to \hbox{the scope of this manuscript}.

\vspace*{\paragraphspacing}
\subparagraph*{Contraction-Based.}
Keuper~et~al.~\cite{keuper2015efficient} propose an algorithm named Greedy Additive Edge Contraction (\texttt{GAEC}).
\texttt{GAEC} is an adaptation of greedy agglomeration where the criterion to evaluate the strength of interactions between clusters is the summed weight of the edges shared by them. 
Bailoni~et~al.~\cite{bailoni2019generalized} propose \texttt{GASP}, a framework for hierarchical agglomerative clustering on weighted signed graphs. 
The authors prove that their framework is a generalization to many existing clustering algorithms such as \texttt{GAEC}~\cite{keuper2015efficient} and introduce new algorithms based on yet unexplored special instances of their framework, such as \texttt{HCC-Sum}.
\texttt{GASP} is a bottom-up approach where nodes are initially assigned to their own clusters, which are then iteratively merged in a pair-wise fashion. 
The framework should be configured to use a specific criterion to evaluate the interaction between a pair of clusters.
Some of these criteria are the (average) weighted edge-cut between the clusters and the weight of their shared edge with highest absolute weight.
The unconstrained variety of \texttt{GASP} starts by merging clusters with the largest positive interaction and stops once the remaining clusters share only negative interactions. 
The constrained variety of \texttt{GASP} introduces cannot-merge constraints between pairs of clusters and terminates when all the remaining clusters are constrained against their neighbors.
This approach greedily selects the pair of clusters with highest absolute interaction, which can be either positive or negative. 
When it is positive, the clusters are merged, otherwise a constraint is added to prohibit the clusters from being merged until the end of the algorithm.
Experimentally, the best algorithms contemplated by \texttt{GASP} with respect to the minimization of edge-cut are \texttt{GAEC}~\cite{keuper2015efficient} and \texttt{HCC-Sum}.
Similarly to \texttt{GAEC} and \texttt{HCC-Sum}, our multilevel algorithm also constructs a clustering based on successive contractions of the graph.
Differently than their approach, our algorithm includes local search methods during the uncoarsening of the graph as well as an extra coarsening-uncoarsening cycle \hbox{where the clustering is further refined}.
Levinkov \etal proposed an extension to \texttt{GAEC}, called \texttt{GAEC+KLj}, and apply Kernighan-Lin based local search after computing an initial solution using \texttt{GAEC}~\cite{10.1007/978-3-319-66709-6_9}.

Hua~et~al.~\cite{HuaRandomWalk2020} propose \texttt{FCSG}, an algorithm for signed graph clustering which combines a Random Walk Gap (RWG) mechanism with a greedy shrinking method.
The RWG mechanism makes a random walk on a version of the graph where negative edges are removed and another random walk on a version of the graph where negative edges are made positive. 
Using the two random walk graphs that are constructed, the gap between the two graphs is calculated to give information on the natural clustering structure of the signed graph. 
This information is then used to build a new signed graph whose edge weights more accurately reflect the natural clustering structure of the input signed graph.
A clustering is then computed on this new signed graph based on a greedy shrinking approach.
Differently than \texttt{FCSG}, our multilevel algorithm does not use random walks and finds a clustering directly on the input signed graph.
Although the coarsening phase of our multilevel algorithm is also based on the gradual contraction of clusters, our algorithm uses this clustering construction approach just to produce an initial solution, which is then refined by \hbox{powerful local search methods}.

Beier \etal \cite{DBLP:conf/cvpr/BeierKKKH14} introduce Cut, Glue \& Cut (CGC), a fast, approximate solver for the NP-hard multi-cut partitioning problem aimed at improving unsupervised image segmentation. Their method iteratively refines the segmentation by first partitioning a graph into smaller regions (cut phase), then exploring alternative segmentations of these regions (glue \& cut phase), efficiently approximating solutions even for non-planar problems.
Levinkov \etal \cite{10.1007/978-3-319-66709-6_9} provide an empirical comparison of local search algorithms for correlation clustering (including GAEC, CGS and more algorithms proposed in \cite{DBLP:conf/cvpr/BeierHK15,DBLP:conf/cvpr/BeierKKKH14,DBLP:conf/iccv/KeuperLBLBA15}), demonstrating their effectiveness in a variety of tasks such as image segmentation and social network analysis. The combination of \texttt{GAEC} with Kernighan-Lin based local search stood out as the producing the best solutions. Hence, we compare our algorithms against this algorithm in our experimental~section.

Recently, shared-memory algorithms for correlation clustering have been proposed~\cite{DBLP:journals/pvldb/ShiDELM21}. We also compare against this algorithm in the experimental part of this paper.  %
 Similarly, an algorithm in the massively parallel model has been proposed~\cite{DBLP:conf/icml/Cohen-AddadLMNP21}. However, this algorithm works on complete graphs only.

\vspace*{\paragraphspacing}
\subparagraph*{Evolutionary.}
Che~et~al.~\cite{memeticsiggraphcommdetec2020} propose \texttt{MACD-SN}, an evolutionary algorithm for signed graph clustering optimizing for \emph{signed modularity}. 
In their algorithm, an individual represents a clustering and the \emph{signed modularity}~\cite{gomez2009analysis} metric is used as fitness function. 
An individual is denoted by an $n$-sized string where each node is assigned to a cluster.
Individuals for the initial population are built in two steps.
First, randomly assign all nodes to clusters.
Second, go through nodes in order to assign each node $u$ to its neighboring cluster $V_i$ which minimizes the expression $NID(u,V_i) = |N^-\cap V_i| + |N^+\cap \overline{V_i}|$. 
Classical approaches are implemented for selection, recombination, and mutation. 
Namely, a tournament selection, a randomized two-way crossover recombination, and a uniform random mutation.
The authors also implement a mutation operator which finds the cluster whose nodes contain the highest average value of $NID(u,V_i)$ and then simply assign each of its nodes $u$ to the cluster which minimizes $NID(u,V_i)$.
Finally, a local search algorithm is executed on the best mutated offspring in each generation.
While the operators of \texttt{MACD-SN} are rather simple and traditional, the operators of our evolutionary algorithm are based on a sophisticated multilevel algorithm which we propose in this work. 
However, the approach only scales to networks \hbox{with a few thousand nodes}.

\vspace*{\paragraphspacing}
\subparagraph*{Integer Linear Programming.}
Many integer linear  programming (ILP) formulations were proposed for the signed graph clustering problem to optimize for edge-cut, as we do here.
Gr{\"{o}}tschel~and~Wakabayashi~\cite{grotschel1989cutting} propose an ILP formulation consisting of $\Theta(n^2)$ binary variables representing pairs of nodes and $\Theta(n^3)$ transitivity constraints.
These constraints ensure that, if two nodes $u,v$ are clustered together, then any other node $w$ will be clustered together with $u$ if, and only if, it is also clustered together with $v$.
Dinh~and~Thai~\cite{dinh2015toward} remove a set of redundant constraints from the ILP by Gr{\"{o}}tschel~and~Wakabayashi~\cite{grotschel1989cutting} for a special case of the signed clustering problem and Miyauchi~and~Sukegawa~\cite{miyauchi2015redundant} extend the ILP by Dinh~and~Thai~\cite{dinh2015toward} to signed graph clustering in general.
The formulation by Miyauchi~and~Sukegawa~\cite{miyauchi2015redundant} removes transitivity constraints associated with two or more edges with negative weight, which reduces the number of constraints to $O(n(n^2-m^-))$ constraints.
Miyauchi~et~al.~\cite{miyauchi2018exact} obtain an improved amount $O(nm^+)$ of constraints by running the ILP from~\cite{miyauchi2015redundant} on a new signed graph where artificial edges are introduced between pairs of unconnected nodes from the input graph.
In particular, these artificial edges have a negative weight which is small enough to ensure that every optimal solution in this new graph corresponds to an optimal solution in the input graph.
Recently Koshimura~et~al.~\cite{koshimura2022concise} proposed a further improvement for the ILP by Miyauchi~et~al.~\cite{miyauchi2018exact} which removes around $50\%$ of its constraints.
Although there have been many improved ILP formulations for signed graph clustering in the last years, this approach does not scale well for large instances (in particular for the size of the instances considered in this paper).
In particular, signed graphs with more than a few thousand nodes become intractable in practice, which is the reason why most instances in the works cited above have \hbox{no more than a few hundred nodes}.

\begin{algorithm*}[t]
	\caption{Our multilevel algorithm}
	\label{alg:overall_multilevel_strategy}
	\hspace*{0cm}\textbf{Input} graph $G=(V,E)$ \\
	\hspace*{0cm}\textbf{Output} clustering $\Pi: V \rightarrow \MdN$ 
	\begin{algorithmic}[1]  
		\STATE $G_0 \gets \emptyset$; $G_1 \gets G$; $i \gets 0$ %
		\WHILE[coarsening phase]{$G_i \neq G_{i+1}$}
		\STATE $i \gets i+1$
		\STATE ComputeClustering($G_i$) %
		\STATE $G_{i+1} \gets$ Contract($G_i$)  %
		\ENDWHILE
		\STATE $\Pi_i \gets$ MapNodesToClusters($G_i$) %
		\FOR[uncoarsening phase]{$j\gets i-1,\ldots,1$}
		\STATE $\Pi_j \gets$ RemapToFiner$(\Pi_{j+1})$ %
		\STATE $\Pi_j \gets$ RefineClusteringLabelPropagation$(G_j,\Pi_j)$ %
		\STATE $\Pi_j \gets$ RefineClusteringFM$(G_j,\Pi_j)$ %
		\ENDFOR
		\STATE $\Pi \gets$ GlobalSearch$(\Pi_{1})$ %
	\end{algorithmic}
\end{algorithm*}

\section{Multilevel Signed Graph Clustering}
\label{sec:Multilevel Signed Graph Clustering}

We now describe our novel multilevel algorithm for signed graph clustering and then we discuss \hbox{each of its components}.

\subsection{Overall Multilevel Strategy.}
\label{subsec:Overall Multilevel Strategy}

Our multilevel approach for graph partitioning differs from others in that it lacks a separate algorithm for initial solution computation. 
Instead, our scheme starts with coarsening to produce a hierarchy of coarser graphs. 
Upon completion, the coarsest graph is assigned individual clusters for each node, resulting in an initial clustering. 
The uncoarsening phase maps the current best clustering all the way back to the original graph while running local search approaches on each level. 
Two consecutive cycles of this multilevel process are performed. %
In the second cycle, the previously obtained clustering is forced as the initial solution on the coarsest level, to be further refined during uncoarsening.
Our overall multilevel algorithm \hbox{is outlined in Algorithm~\ref{alg:overall_multilevel_strategy}}.

\subparagraph*{Coarsening.}
\label{subsec:Coarsening}
As shown in Algorithm~\ref{alg:overall_multilevel_strategy}, our \emph{coarsening} phase consists of two consecutive steps which are iteratively repeated on the current coarsest graph until no more contraction can be performed without increasing the overall edge-cut value. 
In the first step, a graph clustering is computed.
In the second step, each cluster contained in this clustering is contracted \hbox{in order to produce a coarser graph}. 
To be self-contained, we briefly outline our clustering algorithm, which is based on \emph{label propagation}~\cite{labelpropagationclustering}.
We chose label propagation here, contracting clusters decreases graph size significantly in multi-level algorithms.
The algorithm starts with each node in its own cluster, i.e., its initial cluster ID is set to its node ID.
The algorithm then works in rounds.  
In each round, all nodes are traversed, and each node $u$ is moved to the cluster with the largest connection weight with $u$, i.e., it is only moved to the cluster $V_i$ with maximum connection weight $\omega(\{(u, v) \mid v \in N(u) \cap V_i \})$ if it is strictly positive. Note that this equation is also evaluated for the block the node is currently contained in, i.e.~if the node has the strongest connection towards the block it is currently contained in, it stays there. Also note that if the cluster of the current node is a singleton the value of the equation for its block equals zero. Lastly, if the maximum connection strength is negative, then the node is placed in its own cluster as a singleton. 
Ties are broken randomly. 
This greedy approach ensures that the edge-cut is monotonically decreased. 
In a graph with $n$ nodes and $m$ edges, one round of label propagation can be implemented in $O(n+m)$ time. 
The algorithm is repeated at most $L$ times, where $L$ is a tuning parameter. 
However, the algorithm is not too sensitive about the concrete choice of the parameter.
The intuition of the overall procedure is to force positive edges inside clusters and negative edges between clusters, which is \hbox{favorable for the edge-cut objective}.

After label propagation, each cluster is replaced by a single node, 
creating a hierarchy of coarser graphs. 
The contraction process, defined in Section~\ref{sec:preliminaries}, ensures a clustering of the coarse graph corresponds to the same edge-cut in finer graphs. 
This aggressive graph-contraction strategy allows us to shrink irregular networks, and the coarsening stops when no further contraction can be done without increasing the edge-cut. 
The final output is the coarsest graph, which implies an initial clustering of assigning each node \hbox{to its own cluster}.

\subparagraph{Uncoarsening.}
\label{subsec:Uncoarsening}
The \emph{uncoarsening} phase is executed directly after the coarsening phase has been completed and every node of the coarsest graph has been assigned to its own cluster.
As shown in Algorithm~\ref{alg:overall_multilevel_strategy}, our uncoarsening phase consists of two consecutive steps which are iteratively repeated for each graph in our contraction hierarchy. %
In the first step, the current clustering is mapped to the graph contained in the current level of the contraction hierarchy.
In the second step, we apply a sequence of local refinement methods to optimize \hbox{our clustering by moving nodes between clusters}.
At each level of uncoarsening, we remap the best clustering to the next finer graph using our contraction approach, maintaining edge-cut. %
We apply two refinement methods: 
First, a \emph{label propagation} refinement randomly visits all nodes and greedily moves each one to the cluster where edge-cut is minimized.
The only difference to coarsening is that we start from the current given clustering and not from the singleton clustering. 
Second, a \emph{Fiduccia-Mattheyses} (FM)~\cite{fiduccia1982lth} refinement greedily goes through the boundary nodes trying to relocate them with a more global perspective in order to improve solution quality. Label propagation is applied first as this helps to quickly get to a good solution, and FM search is applied afterwards to escape local optima.
Both methods are based on the concept of \emph{Gain}, defined as the decrease \hbox{in edge-cut caused by node movement}.

\begin{algorithm*}[t]
	\caption{Our distributed memetic algorithm}
	\label{alg:overall_evolutionary_strategy}
	\hspace{0cm}\textbf{Input} graph $G=(V,E)$ \\
	\hspace{0cm}\textbf{Output} clustering $\Pi: V \rightarrow \MdN$ 
	\begin{algorithmic}[1]  
		\STATE $P \gets \{\Pi_1, \ldots, \Pi_\alpha\}$ \COMMENT{generate $\alpha$-sized population using diff.~random seeds}
                \STATE \COMMENT{for multilevel algorithm. Each $\Pi_i$ is a clustering.}
		\WHILE{running time $< t_f$}
		\STATE $i \gets$ Random$([0,1])$ %
		\IF[with probability $1-\beta$]{$i > \beta$}
		\STATE $\Pi_{a}, \Pi_{b} \gets $ Selection($P$) \COMMENT{select individuals}
		\STATE $\Pi_c \gets $ Recombination($P$)  %
		\ELSE[with probability $\beta$]
		\STATE $\Pi_c \gets $ Mutation($\Pi_a$) \COMMENT{mutate individual}
		\ENDIF
		\STATE $P \gets $ Replacement($P,\Pi_c$)  \COMMENT{replacement}
		\STATE CommunicateBest($P$)  %
		\ENDWHILE
		
		\STATE $\Pi \gets $ BestIndividualOverall() %
	\end{algorithmic}
\end{algorithm*}

\vspace*{\paragraphspacing}
\subparagraph*{FM Refinement.}
\label{subsec:K-Way_FM_Refinement}

Our FM refinement is an adapted version of the FM algorithm~\cite{fiduccia1982lth} to move nodes between clusters.
We use a gain-based priority queue initialized with the \emph{complete} partition boundary.
Initially all nodes are unmarked. Only unmarked nodes can be in the priority queue. Thus in each round each node is moved at most once.
We repeatedly move the highest-gain node to its best neighboring cluster, updating the queue by adding its unmarked neighbors not yet moved. 
Once a node is moved it is marked. 
The refinement escapes local optima by allowing negative-Gain movements and rolling back to the lowest cut after the search has concluded.
The search concludes once the queue becomes empty, indicating that each node has been processed once, or after fifteen consecutive moves that yield no positive gain. We use fifteen here as this is default for FM-based local search algorithms, such as those implemented in Metis~\cite{karypis1998fast}, which is designed to address the challenge of balanced graph partitioning.

\subparagraph*{Global Search.}
\label{subsec:Global Search}

After the multilevel cycle described in Algorithm~\ref{alg:overall_multilevel_strategy},  an extra multilevel cycle is run which uses the output clustering of the first cycle as the initial solution.
This global search strategy was introduced in \cite{walshaw2004multilevel} and has been successfully used for many optimization problems.
The core difference to the first multilevel cycle is that the previously computed clustering is forced to be the initial solution in the extra multilevel cycle. 
This is achieved by blocking cut edges from contraction and stopping coarsening when the coarser graph equals the quotient graph of the clustering. 
More precisely, 
we do this by modifying the label propagation algorithm during coarsening such that each cluster of the computed clustering is a subset of a cluster of the input clustering from the first cycle. In other words, each cluster only contains nodes of one unique clustering of the input clustering. 
Hence, when contracting the clustering, every cut edge of the input clustering will remain.
This approach is extended in Section~\ref{sec:Distributed Evolutionary Signed Graph Clustering} to \hbox{form the operators of our memetic algorithm}.

\section{Distributed Memetic Signed Graph Clustering}
\label{sec:Distributed Evolutionary Signed Graph Clustering}

Our distributed memetic algorithm is summarized in Algorithm~\ref{alg:overall_evolutionary_strategy}. We first explain how the algorithm runs sequentially and then explain our coarse-grained parallelization. We start by giving an overview of our algorithm.
As a typical memetic approach, our algorithm is a population-based heuristic. %
Given a signed graph~$G$, we define an \emph{individual} as a particular clustering on $G$.
First, our algorithm initializes a population~$P$ of $\alpha$ individuals using different random seeds.
Then the algorithm proceeds in rounds. In each round the population evolves via either \emph{mutation} (with probability $\beta=10\%$) or \emph{recombination} (with probability $1-\beta=90\%$). 
The process continues until a desired time limit~$t_f$ is reached.
The \emph{recombination} procedure is executed on two \emph{good} individuals selected via \emph{tournament} selection and then used as \emph{parents} to produce \emph{offspring} which partially inherits their characteristics.
The \emph{mutation} procedure is run on a randomly-picked individual which is used as reference for building another individual from scratch. 
The next phase is called \emph{replacement}.
Newly computed individuals are inserted into the population and the worst ones are \emph{evicted} to maintain $\alpha$ population size.
Since our algorithm generates only one individual per round it falls into the category of \emph{steady-state} evolutionary algorithms~\cite{evolutionary_book}.

\subparagraph{Notation, Population, and Fitness.}
\label{subsec:Notation, Population, and Fitness}

We define an individual as a clustering $\Pi$ on the graph $G$ and represent it by its set of cut edges.
Note that some sets of edges cannot be the set of cut edges of any feasible clustering, e.g., a single edge of a cycle.
However, feasibility is \emph{implicitly} ensured throughout the algorithm, as will become clear later.
As a consequence, we do not need any penalty function, hence the \emph{fitness} function of an individual is its \emph{edge-cut}, which is the objective to be minimized.
The initial population is built by running our multilevel algorithm (Algorithm~\ref{alg:overall_multilevel_strategy}) from scratch $\alpha$ times, where $\alpha \in [3, 100]$ is determined such that roughly $10\%$ of the allowed time limit $t_f$ is spent \hbox{on the construction of the initial population}. 
More precisely, our algorithm performs one multilevel clustering step and measures the time $t$ spent. We then
choose $\alpha$ such that the time for creating $\alpha$ partitions is approximately $10\%\cdot t_\text{total}$ where $t_\text{total}$ is the total running time that the algorithm is given to produce a clustering of the graph.

\subparagraph*{Recombination.}
\label{subsec:Recombination}

\label{subsec:Evolutionary Strategy}
\begin{figure*}[t!]
	\centering
	\includegraphics[width=.6\textwidth]{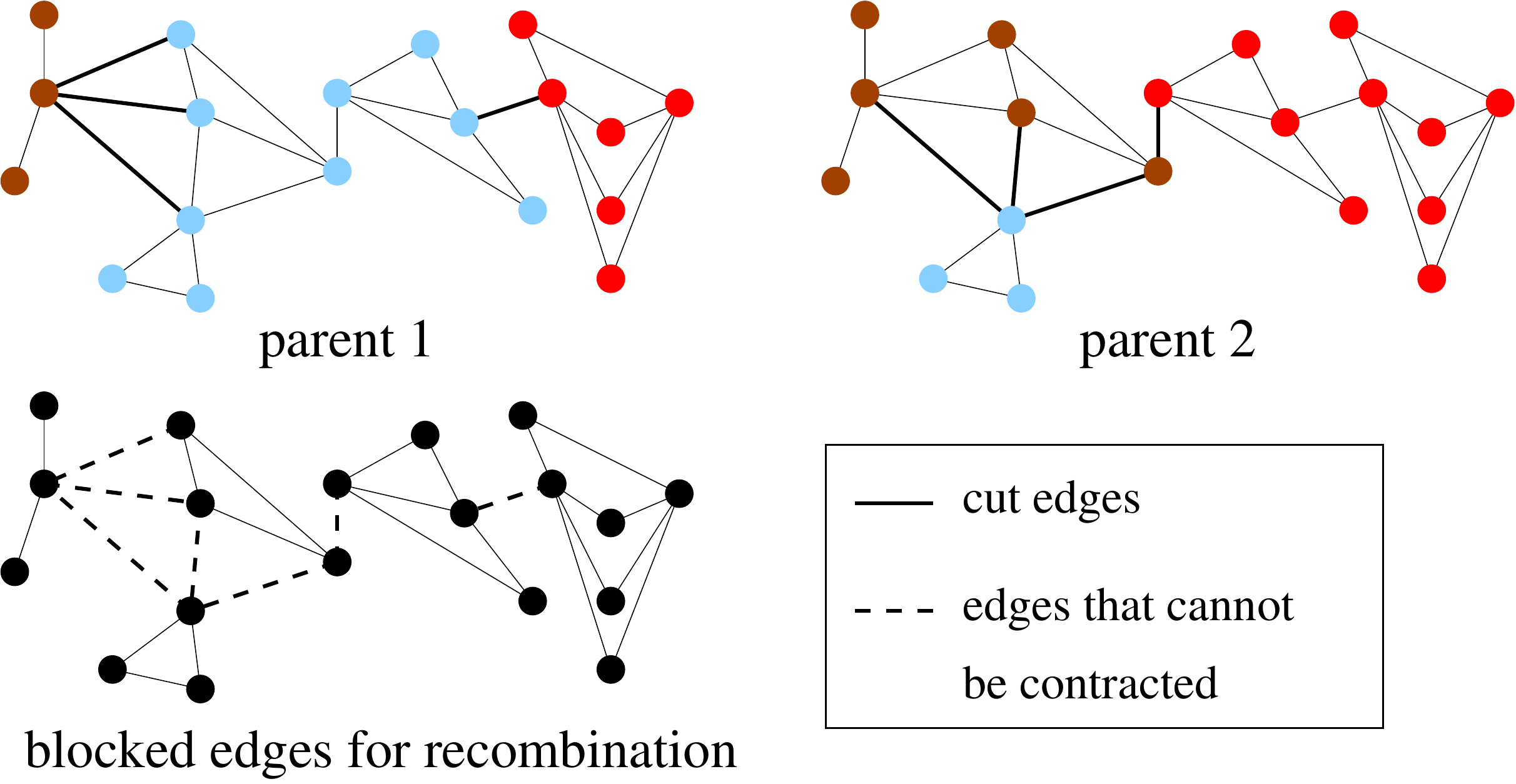}
	\caption{In the recombination, the cut edges of both parents cannot be contracted during the first multilevel cycle to build an offspring.}
	\label{fig:recombination}
\end{figure*}

Our recombination algorithm works as follows. First, we select two \emph{good} individuals, $\Pi_a$ and $\Pi_b$, from the population to serve as parents using a \emph{tournament selection}~\cite{tournament_selection}, i.e., two distinct individuals are drawn and the one with best edge-cut is selected to be a parent. 
To ensure distinct parents, the second parent may be the loser of the tournament if the winner is already the first parent. 
To create an offspring, we run Algorithm~\ref{alg:overall_multilevel_strategy} on $G$ and block the cut edges of $\Pi_a$ and $\Pi_b$ from contracting during the first multilevel cycle. 
More precisely, we do this by modifying the label propagation algorithm during coarsening such that each cluster of the computed clustering is a subset of any cluster of both of the input clusterings. 
That is a cluster found by label propagation can only be found within a cluster of the clustering induced by the connected components of the graph where all cut edges of $\Pi_a$ and $\Pi_b$ have been removed. 
Figure~\ref{fig:recombination} demonstrates which edges are blocked for a particular instance.
This leads to a coarsening phase that stops when no further contraction is possible unless the edge-cut increases or blocked edges are contracted. 
We then choose the best edge-cut clustering among three options: the two parents and a clustering where each node of the coarsest graph is assigned its own cluster. 
Our algorithm then uncoarsens this new clustering, applies local searches, and further improves the clustering with the global search in Figure~\ref{fig:MSGC_second_cycle}. 
This approach directly ensures non-increasing edge-cut compared to $\Pi_a$ and $\Pi_b$ and combines the characteristics of both parents, \hbox{as expected from recombination}.

\subparagraph*{Mutation.}
\label{subsec:Mutation}

For mutation, we randomly select an individual $\Pi_a$ from the population and create a new individual by running Algorithm~\ref{alg:overall_multilevel_strategy} on $G$ while blocking $\Pi_a$'s cut edges from contracting only in the \emph{first level} of the first multilevel cycle. 
The initial solution obtained at the finer level of this multilevel cycle can differ from $\Pi_a$ but inherits some of its characteristics due to this edge-blocking constraint. 
The new individual is obtained by optimizing this initial solution during uncoarsening as well as the global search cycle. 
This operator does not guarantee non-increasing edge-cut compared to $\Pi_a$, although it increases population variability and optimizes \hbox{the new individual as much as possible}.

\subparagraph*{Replacement.}
\label{subsec:Replacement}

Our replacement strategy inserts the newly generated individual $\Pi_c$ from mutation or recombination into the population $P$ and removes an individual to keep only~$\alpha$ individuals alive. 
We first check if $\Pi_c$ has worse edge-cut than all individuals in $P$. 
If so, we do not insert $\Pi_c$ into $P$. 
Otherwise, we remove the individual in $P$ most similar to $\Pi_c$, where a high similarity is defined as a low-cardinality symmetric difference between the two individuals' edge sets. 
This approach helps maintain high diversity in $P$, which is \hbox{beneficial for the evolutionary process}.

\subparagraph*{Parallelization.}
\label{subsec:Parallelization}

Our parallelization works as follows. Each processor performs  the same steps outlined in Algorithm~\ref{alg:overall_evolutionary_strategy} (using different random seeds). 
Communication is done through a variation of the \emph{randomized rumor spreading} protocol~\cite{randomgossip}. 
The communication is organized in rounds, in which each process tries to send and receive individuals. 
In each round, a process sends the best individual from its local population to a randomly selected process it has not yet sent this individual to. 
Afterwards, it tries to receive incoming individuals and inserts them into its local population using the described replacement strategy. 
The overall Algorithm~\ref{alg:overall_evolutionary_strategy} is performed asynchronously, i.e., without global synchronization. 

\nprounddigits{0}
\begin{table}[t!]
	\centering
	\scriptsize
	\footnotesize
	\setlength{\tabcolsep}{15pt}
        \centering
	\begin{tabular}{l@{\hskip 20pt}r@{\hskip 20pt}r@{\hskip 20pt}r@{\hskip 10pt}}
		\toprule	
		Graph          & Nodes  & Edges  & Source \\ 
                \midrule
                \Id{knott3D-32} &\numprint{651}&\numprint{3939} &\cite{DBLP:journals/corr/abs-2202-03574} \\
                \Id{knott3D-72} &\numprint{4228}&\numprint{26278} & \cite{DBLP:journals/corr/abs-2202-03574}\\
                \Id{knott3D-96} &\numprint{15393}&\numprint{97275}& \cite{DBLP:journals/corr/abs-2202-03574}\\
                \midrule
                \Id{img167083} &\numprint{2904}& \numprint{8409}& \cite{DBLP:journals/corr/abs-2202-03574}\\
                \Id{img175032} &\numprint{3764}&\numprint{10898}&  \cite{DBLP:journals/corr/abs-2202-03574}\\
                \Id{img236037} &\numprint{2952}&\numprint{8559}&  \cite{DBLP:journals/corr/abs-2202-03574}\\
        \midrule
		\Id{bitcoinalpha}   & \numprint{3783}   & \numprint{14081}  & \cite{snap}\\
		\Id{bitcoinotc}     & \numprint{5881}   & \numprint{21434}  & \cite{snap}\\
		\Id{elec}           & \numprint{7118}   & \numprint{100355} & \cite{konect}\\
		\Id{chess}          & \numprint{7301}   & \numprint{32650}  & \cite{konect}\\
		\Id{slashdot081106} & \numprint{77357}  & \numprint{466666} & \cite{snap}\\
		\Id{slashdot-zoo}   & \numprint{79116}  & \numprint{465840} &\cite{snap} \\
		\Id{slashdot090216} & \numprint{81871}  & \numprint{495666} & \cite{snap}\\
		\Id{slashdot090221} & \numprint{82144}  & \numprint{498532} & \cite{snap}\\
		\Id{wikiconflict}   & \numprint{118100} & \numprint{1461058}& \cite{konect}\\
		\Id{epinions}       & \numprint{131828} & \numprint{708507} & \cite{snap}\\
		\Id{wikisigned-k2}  & \numprint{138592} & \numprint{712337} & \cite{konect}\\ 
                \midrule
                \Id{cityscape7}  & \numprint{2097152} & \numprint{9332736}&\cite{DBLP:journals/corr/abs-2202-03574}\\
                \Id{cityscape34} & \numprint{2097152}&\numprint{9332736}&\cite{DBLP:journals/corr/abs-2202-03574}\\
                \Id{cityscape52} & \numprint{2097152}&\numprint{9332736}&\cite{DBLP:journals/corr/abs-2202-03574}\\
                \bottomrule
	\end{tabular}
	\caption{Properties of real-world signed graphs.}
        \vspace*{-.5cm}
	\label{tab:graphs}
\end{table}

\nprounddigits{0}
\begin{table*}[t!]

                \centering
\begin{tabular}{lrrr||rrrr}
\toprule
Graph & \texttt{SCML}  & \texttt{SCMLEvo} & \texttt{SCMLEvoPar} & \texttt{GAEC} & \texttt{GAEC+KL} & \texttt{HCC} & \texttt{ParCCML}\\
              \midrule
\Id{knott3d032}     & \numprint{-1882}    & \textbf{\numprint{-1884}}  & \textbf{\numprint{-1884}}    & \numprint{-1879}    & \numprint{-1883}           & \numprint{-1868}    &\numprint{-1876}        \\
\Id{knott3d072}     & \numprint{-11581}   & \textbf{\numprint{-11618}} & \textbf{\numprint{-11618}}   & \numprint{-11559}   & \numprint{-11609}          & \numprint{-11521}   &\numprint{-11499}    \\
\Id{knott3d096}     & \numprint{-31630}   & \textbf{\numprint{-31754}} & \textbf{\numprint{-31754}}   & \numprint{-31557}   & \numprint{-31711}          & \numprint{-31442}   &\numprint{-31526}    \\
\midrule                                                                                                                                                                                            
\Id{img167083}      & \numprint{-3482}    & \textbf{\numprint{-3510}}  & \textbf{\numprint{-3510}}    & \numprint{-3478}    & \numprint{-3496}           & \numprint{-3464}    &\numprint{-3482}     \\
\Id{img175032}      & \numprint{-4822}    & \numprint{-4858}           & \textbf{\numprint{-4859}}    & \numprint{-4759}    & \numprint{-4817}           & \numprint{-4733}    &\numprint{-4792}       \\
\Id{img236037}      & \numprint{-4289}    & \textbf{\numprint{-4310}}  & \textbf{\numprint{-4310}}    & \numprint{-4251}    & \numprint{-4293}           & \numprint{-4232}    &\numprint{-4258}     \\
\midrule                                                                                                                                                                                            
\Id{bitcoinalpha}   & \numprint{-5477}    & \numprint{-5534}           & \numprint{-5534}             & \numprint{-5531}    & \textbf{\numprint{-5561}}  & \numprint{-5476}    &\numprint{-5558}        \\
\Id{bitcoinotc}     & \numprint{-20236}   & \numprint{-20425}          & \numprint{-20429}            & \numprint{-20395}   & \textbf{\numprint{-20433}} & \numprint{-20327}   &\numprint{-20328}      \\
\Id{elec}           & \numprint{-7706}    & \textbf{\numprint{-7735}}  & \textbf{\numprint{-7735}}    & \numprint{-7722}    & \numprint{-7724}           & \numprint{-7722}    &\numprint{-7706}        \\
\Id{chess}          & \numprint{-4421}    & \numprint{-4775}           & \textbf{\numprint{-4802}}    & \numprint{-4346}    & \numprint{-4496}           & \numprint{-4304}    &\numprint{-4095}        \\
\Id{slashdot081106} & \numprint{-49004}   & \numprint{-49791}          & \textbf{\numprint{-49806}}   & \numprint{-48528}   & \numprint{-49042}          & \numprint{-48501}   &\numprint{-49177}      \\
\Id{slashdot-zoo}   & \numprint{-53430}   & \numprint{-53696}          & \textbf{\numprint{-53705}}   & \numprint{-52394}   & \numprint{-52975}          & \numprint{-52347}   &\numprint{-53006}      \\
\Id{slashdot090216} & \numprint{-50276}   & \numprint{-50655}          & \textbf{\numprint{-50669}}   & \numprint{-49875}   & \numprint{-50293}          & \numprint{-49870}   &\numprint{-49885}      \\
\Id{slashdot090221} & \numprint{-50274}   & \numprint{-50914}          & \textbf{\numprint{-50926}}   & \numprint{-49822}   & \numprint{-50329}          & \numprint{-49821}   &\numprint{-50180}      \\
\Id{wikiconflict}   & \numprint{-2167064} & \numprint{-2167366}        & \textbf{\numprint{-2167377}} & \numprint{-2166609} & \numprint{-2167259}        & \numprint{-2166610} &\numprint{-2166505}    \\
\Id{epinions}       & \numprint{-70131}   & \numprint{-70316}          & \textbf{\numprint{-70321}}   & \numprint{-69426}   & \numprint{-70205}          & \numprint{-68769}   &\numprint{-69558}       \\
\Id{wikisigned-k2}  & \numprint{-41236}   & \numprint{-41526}          & \textbf{\numprint{-41562}}   & \numprint{-41465}   & \numprint{-41482}          & \numprint{-41398}   &\numprint{-41099}       \\
\midrule                                                                                                                                                                                            
\Id{cityscape7}     & \numprint{-435564}  & \numprint{-440469}         & \textbf{\numprint{-441318}}  & \numprint{-365560}  & $\times$                   & \numprint{-351701}  &\numprint{-377516}    \\
\Id{cityscape34}    & \numprint{-187085}  & \numprint{-188787}         & \textbf{\numprint{-188979}}  & \numprint{-112591}  & \numprint{-186589}         & \numprint{-81550}   &\numprint{-140412}    \\
\Id{cityscape52}    & \numprint{-293378}  & \numprint{-296039}         & \textbf{\numprint{-296291}}  & \numprint{-257493}  & $\times$                   & \numprint{-248931}  &\numprint{-225763}     \\

                \bottomrule
	\end{tabular}
\caption{Minimum edge-cuts achieved by various algorithms over ten repetitions. Instances marked with $\times$ did not finish within a 72h time limit. \texttt{SCMLEvo} and \texttt{SCMLEvoPar} both ran for two minutes. Lowest values are highlighted in \textbf{bold} font.}

                \label{tab:results}

        \end{table*}
\vfill \pagebreak
\section{Experimental Evaluation.}
\label{sec:Experimental Evaluation}

\subparagraph*{Methodology.} 
We implemented our algorithms using C++ and compiled them as well as competing C++ codes using gcc 11.4 with full optimization turned on (-O3 flag). 
We have used a machine with an  AMD EPYC 9754 128-Core CPU running at 2.25GHz with 256MB L3 Cache and 768GB of main memory. 
It runs Ubuntu GNU/Linux 22.04 \hbox{and Linux kernel version 5.15.0-102}.

Depending on the focus of the experiment, we measure running time and/or edge-cut.
An alternative metric for edge-cut is the \emph{z\_value}~\cite{HuaRandomWalk2020}, which can be mathematically defined as $1-\frac{edgecut}{\omega(E^{-})}$ where lower values are better.
Unless explicitly mentioned otherwise, we run our experiments ten times on each of the graphs listed Table~\ref{tab:graphs} with different random seeds.
When performing repetitions per algorithm and graph using different random seeds for initialization, we compute the minimum value of the computed objective functions 
 and running time per instance.
When further averaging over multiple instances, we use the geometric mean.
Since all values of edge-cut computed in our experiments are negative, we define geometric mean as the geometric mean of its absolute value multiplied afterwards by $-1$.

\nprounddigits{3}
\begin{table*}[t]
\setlength{\tabcolsep}{5pt}
\begin{center}
\begin{tabular}{lr||rrrr|||rr}
\toprule

Graph                              & \texttt{SCML}                      & \texttt{GAEC}       & \texttt{GAEC+KLj}       & \texttt{HCC}   & \texttt{ParCCML} & Speedup & \multicolumn{1}{c}{Speedup} \\ 
&                     &      &   &  & &\texttt{SCMLEvo}  & \multicolumn{1}{c}{\texttt{SCMLEvoPar}} \\ 
&                     &      &   &  & & \multicolumn{2}{c}{over}   \\ 
&                     &      &   &  & &\multicolumn{2}{c}{\texttt{GAEC+KLj}}  \\ 
\midrule
\Id{knott3D-32}     & \numprint{0.00374284}          & \textbf{\numprint{0.00180883}} & \numprint{0.0144408} & \numprint{0.002037}           &\numprint{0.0449857}& 0.05    & 2.34   \\
\Id{knott3D-72}     & \numprint{0.025829}            & \textbf{\numprint{0.0126374}}  & \numprint{0.21087}   & \numprint{0.0148045}          &\numprint{0.0761915}& 0.07    & 0.15   \\
\Id{knott3D-96}     & \numprint{0.0878565}           & \numprint{0.0697913}           & \numprint{10.3802}   & \textbf{\numprint{0.0661762}} &\numprint{0.117145} & 1.02    & 1.77  \\
\midrule                                                                                                                                                          
\Id{img167083}      & \numprint{0.00864148}          & \textbf{\numprint{0.00509086}} & \numprint{0.0284701} & \numprint{0.00589583}         &\numprint{0.0547081}& 0.02    & 0.04 \\
\Id{img175032}      & \numprint{0.0153701}           & \textbf{\numprint{0.00496209}} & \numprint{0.0330856} & \numprint{0.00614238}         &\numprint{0.0586023}& 0.10    & 2.65 \\
\Id{img236037}      & \numprint{0.0109259}           & \textbf{\numprint{0.00411868}} & \numprint{0.0195697} & \numprint{0.00546618}         &\numprint{0.0535833}& 0.02    & 2.18 \\
\midrule                                                                                                                                                          
\Id{bitcoinalpha}   & \textbf{\numprint{0.00701392}} & \numprint{0.0126811}           & \numprint{0.291567}  & \numprint{0.00994925}         &\numprint{0.0519295}& $\star$ & $\star$ \\
\Id{bitcoinotc}     & \textbf{\numprint{0.0114954}}  & \numprint{0.0162384}           & \numprint{2.14611}   & \numprint{0.0178193}          &\numprint{0.0603764}& $\star$ & $\star$ \\
\Id{chess}          & \numprint{0.0596461}           & \textbf{\numprint{0.0223447}}  & \numprint{3.65732}   & \numprint{0.024154}           &\numprint{0.0698113}& 0.6     & 2.33\\
\Id{elec}           & \textbf{\numprint{0.0339655}}  & \numprint{0.0511454}           & \numprint{9.32165}   & \numprint{0.0687744}          &\numprint{0.0816346}& 2.08    & 5.27 \\
\Id{slashdot-zoo}   & \textbf{\numprint{0.493301}}   & \numprint{0.936291}            & 5\ 523.89            & \numprint{1.06536}            &\numprint{0.28415}  & 9919    & 13028  \\
\Id{slashdot081106} & \textbf{\numprint{0.38685}}    & \numprint{0.985825}            & 4\ 063.36            & \numprint{1.00747}            &\numprint{0.301597} & 3421    & 4815 \\
\Id{slashdot090216} & \textbf{\numprint{0.382348}}   & \numprint{1.22705}             & 4\ 610.00            & \numprint{1.06482}            &\numprint{0.313937} & 2390    & 9302 \\
\Id{slashdot090221} & \textbf{\numprint{0.390044}}   & \numprint{1.21957}             & 5\ 446.96            & \numprint{1.19687}            &\numprint{0.307433} & 4198    & 12321 \\
\Id{wikiconflict}   & \numprint{2.12}                & \textbf{\numprint{1.71417}}    & 11\ 135.00           & \numprint{2.10525}            &\numprint{0.542221} & 342     & 1192\\
\Id{epinions}       & \textbf{\numprint{0.436731}}   & \numprint{2.05957}             & 4\ 386,99            & \numprint{1.95548}            &\numprint{0.417049} & 176     & 476  \\
\Id{wikisigned-k2}  & \textbf{\numprint{0.605078}}   & \numprint{2.0575}              & 3\ 265.75            & \numprint{1.95867}            &\numprint{0.443065} & 54      & 236 \\
\midrule                                                                                                                                                          
\Id{cityscape7}     & \textbf{10.92}                 & 71.14                          & $\times$             & 70.01                         &\numprint{3.07586}  & $-$     & $-$\\
\Id{cityscape34}    & \textbf{7.98}                  & 66.55                          & 142\ 374.00          & 73.07                         &\numprint{2.69374}  & 7\ 014  & 16\ 807 \\
\Id{cityscape52}    & \textbf{8.83}                  & 74.48                          & $\times$             & 74.29                         &\numprint{2.77051}  & $-$     & $-$ \\
\bottomrule

\end{tabular}
\end{center}
\caption{Running times of various algorithms under consideration in \emph{seconds} (left) as well as time of \texttt{GAEC-KLj} divided by the time to reach result of \texttt{GAEC-KLj} by the memetic algorithms denoted as speedup (right, larger is better). In instances that are marked with $\star$ our algorithm did not reach the result of \texttt{GAEC-KL}. Instances marked with $\times$ did not finish within a 72h time limit. Lowest running are highlighted in \textbf{bold} font. \texttt{SCMLEvo} and \texttt{SCMLEvoPar} both ran for two minutes.}
\setlength{\tabcolsep}{5pt}
\label{tab:resultsrunningtime}

\end{table*}

\vspace*{\paragraphspacing}
\subparagraph*{Instances.}
We use the real-world signed graphs listed in Table~\ref{tab:graphs}.
They were all obtained from the public graph collections SNAP~\cite{snap} and KONECT~\cite{konect} as well as the structured prediction problem archive~\cite{DBLP:journals/corr/abs-2202-03574}. From the latter we choose three randomly selected instances the \Id{cityscape} benchmark~\cite{DBLP:conf/cvpr/CordtsORREBFRS16}, the \Id{knott3D} benchmark~\cite{DBLP:journals/ijcv/KappesAHSNBKKKL15} as well as natural image segmentation benchmark~\cite{DBLP:journals/ijcv/KappesAHSNBKKKL15}.
Before running the experiments, we converted each of the graphs in Table~\ref{tab:graphs} to an undirected signed graph without parallel or self edges. 
In particular, we achieve this by simpl\NEW{y} removing self edges and substituting all parallel and opposite arcs by a single undirected edge whose weight equals the sum of the weights of these arcs.
Positive edges obtain weight one while negative edges obtain weight minus one.

\vspace*{\paragraphspacing}
\subparagraph*{Competitors.}
Based on the experimental study by Levinkov \etal \cite{10.1007/978-3-319-66709-6_9} we compare against \texttt{GAEC}~\cite{keuper2015efficient}, \texttt{GAEC+KLj}~\cite{10.1007/978-3-319-66709-6_9} (best algorithm w.r.t.~edge-cut) and \texttt{HCC-Sum} (\texttt{HCC})~\cite{bailoni2019generalized}. 
For \texttt{GAEC} and \texttt{HCC-Sum} we use the implementations provided in the \texttt{GASP}~framework~\cite{bailoni2019generalized}, which is publicly available. Bailoni~et~al.~\cite{bailoni2019generalized} show that they are the two best algorithms with respect to the minimization of edge-cut -- however the authors do not compare against \texttt{GAEC+KLj}. For \texttt{GAEC+KLj} we use the implementation that has been provided by the authors of \cite{10.1007/978-3-319-66709-6_9} in \url{https://github.com/bjoern-andres/graph}.
We also compare against the recent shared-memory parallel algorithm by Shi \etal~\cite{DBLP:journals/pvldb/ShiDELM21} (\url{https://github.com/google/graph-mining}). We run the algorithm with the default parameters and use multilevel refinement which according to the authors gives better solutions. The algorithm is denoted as \texttt{ParCCML}. Based on the $z\_value$ explicitly reported in~\cite{HuaRandomWalk2020} and our experiments with \texttt{GAEC}, already \texttt{GAEC} without KL local search outperforms \texttt{FCSG}. Moreover, the implementation is not available. Hence, we do not include it here. 
From now on, we refer to our algorithms as follows: \texttt{SCML} represents our multilevel algorithm (\Id{S}igned \Id{C}lustering \Id{M}ulti\Id{L}evel), \texttt{SCMLEvo} represents our memetic algorithm run on a single processor, and \texttt{SCMLEvoPar} represents our memetic algorithm using all cores of our machine.

\subsection{Experimental Comparison.}
In this section, we conduct a series of experiments to assess the extent to which our memetic algorithm surpasses current state-of-the-art methods. 
Table~\ref{tab:results} and Table~\ref{tab:resultsrunningtime} give detailed per instances results in terms of edge-cut achieved as well as running times of different algorithms. 
As \texttt{HCC} is roughly equally fast compared to the \texttt{GAEC} algorithm, and in terms of solution quality is always worse, we do not further discuss it here. 

We begin by comparing our sequential multilevel strategy \texttt{SCML} to the \texttt{GAEC} algorithm. In terms of solution quality our multilevel algorithm is on par with the \texttt{GAEC} algorithm. On 16 out of 20 instances our algorithm computes a slightly better result than the \texttt{GAEC} algorithm. 
However, on some instances improvement in terms of edge-cut is very big. For example, on the \Id{cityscape} instances, \texttt{SCML} computes cuts that are 10\%, 16\% and 40\% lower. 
However, in terms of running time we do see a significant advantage. Overall, our multilevel algorithm \texttt{SCML} is 46\% faster on average. 
While on the very small instances like \Id{knott3D} and \Id{img} our algorithms is slower (up to a factor three), it is much faster on the large instances typically a factor 2 to 3. The largest speedup is observed on the \Id{cityscape} instances. Here, our multilevel algorithm is between a factor 6.46 to 8.43 faster.  
The advantage in speed of our algorithm on large instances stems from the fact that the coarsening process based on contracting clusterings works very well in reducing the instance size. For example, after the \emph{first} contraction step the contracted graph in the hierarchy has only 33.3k nodes for the \Id{cityscape34} instances which is roughly 1.5\% of the original input size.  As another example, after the first contraction step for the \Id{wikisigned-k2} the coarser graph has only 17.2k nodes which is approximately 12.4\% of the original input size.  

The \texttt{GAEC+KLj} algorithm which combines \texttt{GAEC} with Kernighan-Lin local search, is able to improve the results of the \texttt{GAEC} algorithm significantly. 
In all cases in which the algorithm finished within a 72 hour time limit, the algorithm improves the result of the \texttt{GAEC} algorithm. 
On average our \texttt{SCML} algorithms computes solutions that are 0.2\% worse compared to the \texttt{GAEC+KLj} algorithm.
While the \texttt{GAEC+KLj} algorithm is able to improve the results of \texttt{GAEC}, it is not scalable. While it is still able to solve very small instances quickly it has problem as soon as the instances get larger. 
The algorithm is on average a factor 235 and 304 slower than the \texttt{GAEC} algorithm and our \texttt{SCML} algorithm on the instances that it could solve respectively. 
Generally the larger the problem gets, the larger is the gap in running time observed. On the largest instances that \texttt{GAEC+KLj} could solve, \Id{cityscape34}, our \texttt{SCML} algorithm is four orders of magnitude faster, \hbox{while also computing a better result.}
\begin{figure*}[t!]
\begin{center}
        \vspace*{-.5cm}
        \includegraphics[width=5.25cm]{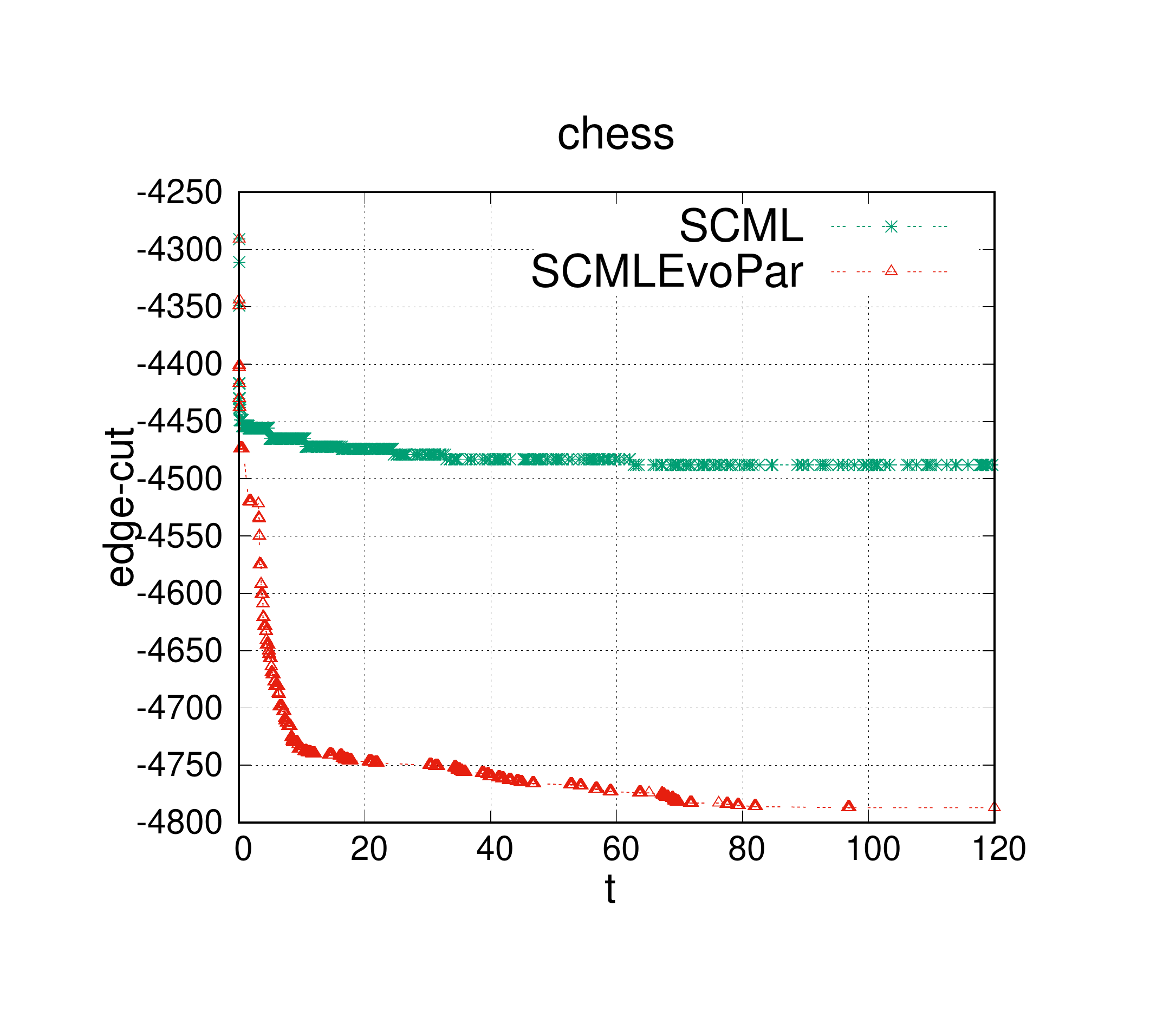}
        \hspace*{-1.25cm}
        \includegraphics[width=5.25cm]{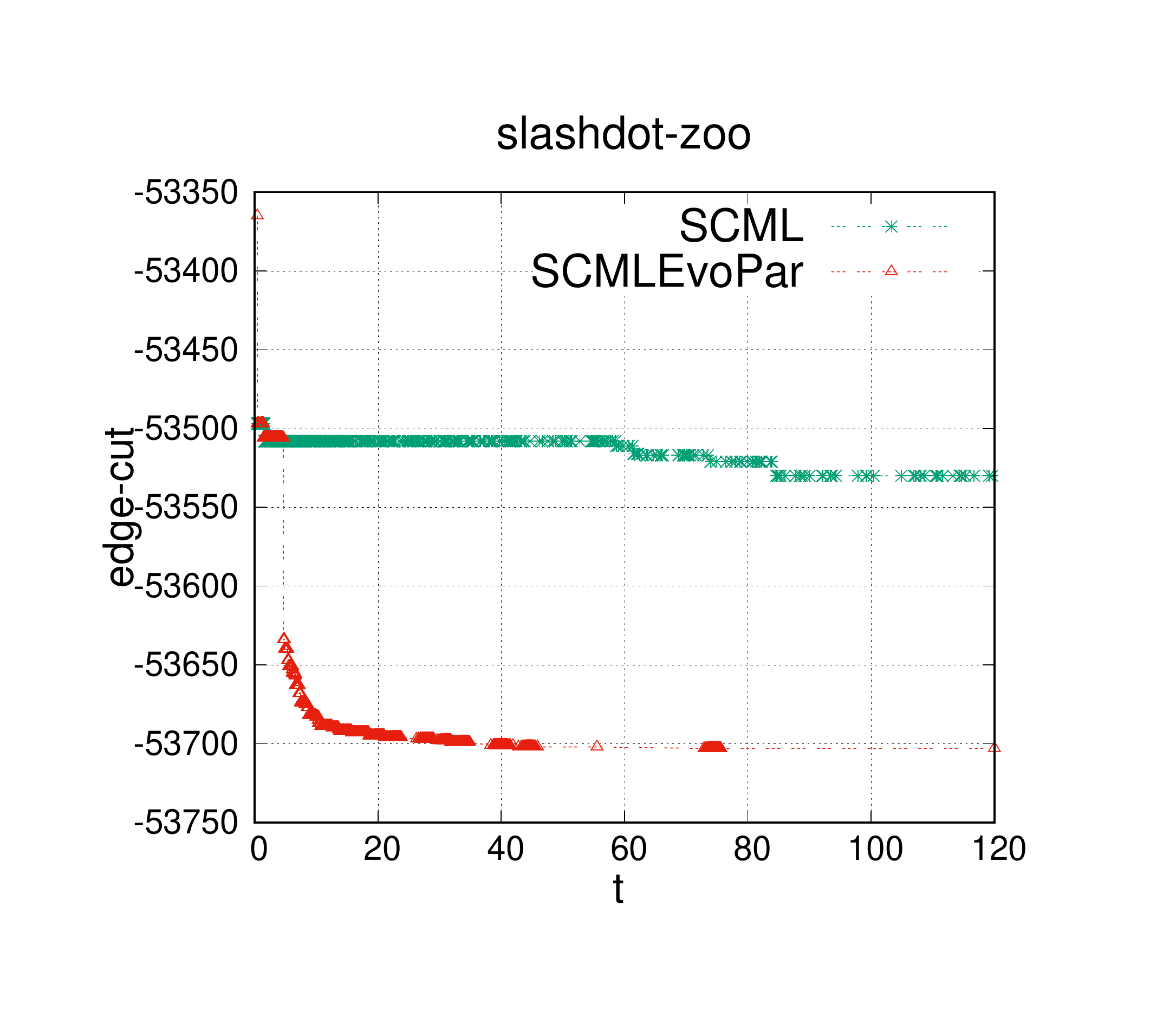}
        \hspace*{-1.25cm}
        \includegraphics[width=5.25cm]{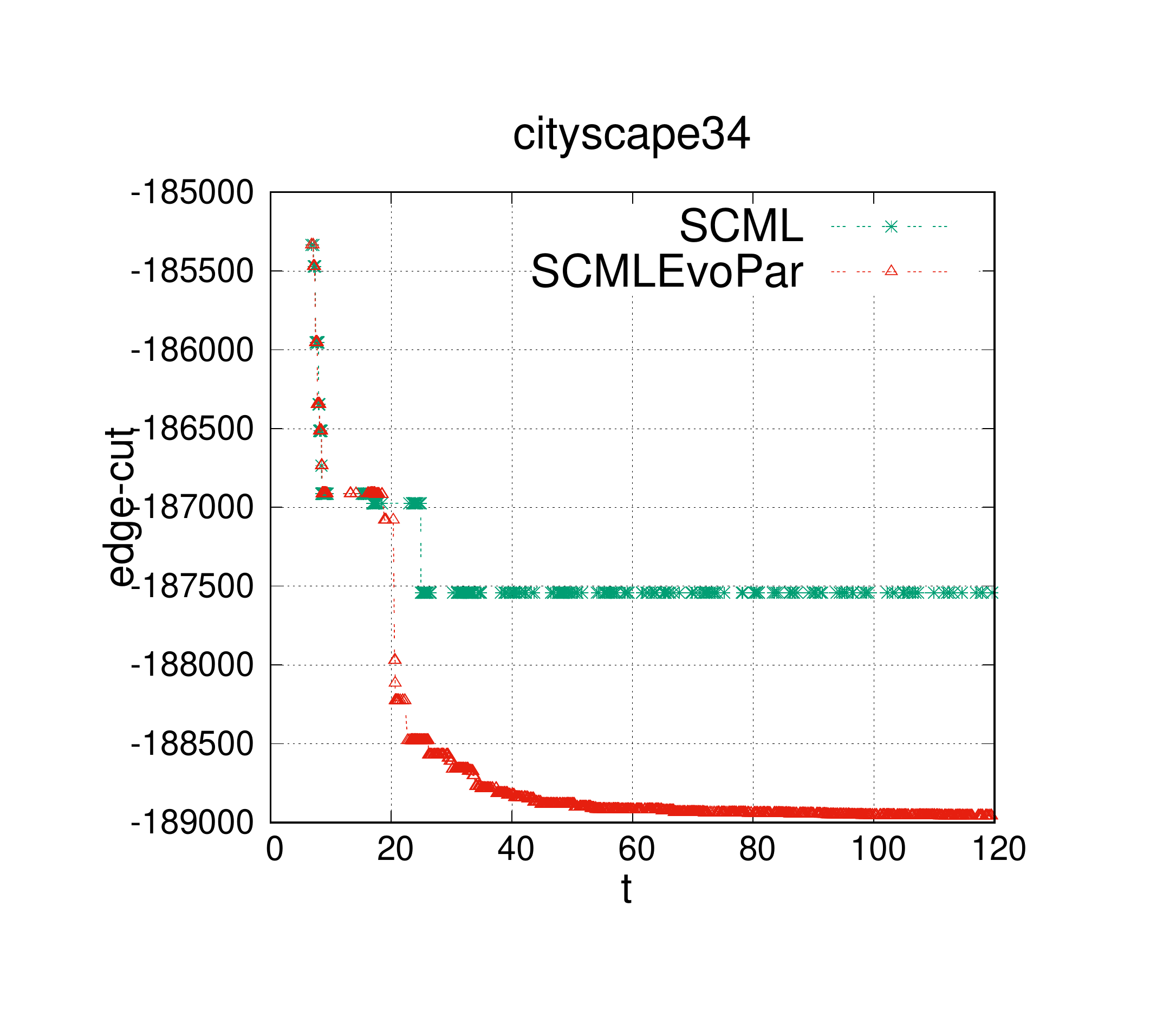}
        \vspace*{-1.25cm}
\end{center}
\caption{Convergence plots for \texttt{SCML} and \texttt{SCMLEvoPar}. Both algorithms run in parallel using all cores of our machine for two minutes to compute a result. The plots show how the best solution computed by the algorithm decreases over time.  }
\label{fig:convergenceplots}

\end{figure*}

We now investigate our sequential and parallel memetic strategies \texttt{SCMLEvo} which runs for 2 minutes sequentially and \texttt{SCMLEvoPar} which runs for 2 minutes on all cores of our machine. 
Already our sequential algorithm \texttt{SCMLEvo} computes better results than \texttt{GAEC+KLj} in all but two cases. These improvements can be big, for example, on the \Id{chess} instance, our algorithm computes a 10\% lower edge-cut.
In order to asses a potential speedup, we measure the time that was needed by our memetic algorithm to reach a clustering with an edge cut similar to that which has been computed by \texttt{GAEC+KLj} algorithm. 
Table~\ref{tab:resultsrunningtime} shows the result. 
We can observe that on small instances there is a significant slow down, i.e.~the~\Id{img} and \Id{knott3D} instances. However, this is somewhat expected as the instances are very small and even the multilevel algorithm itself has been slower as there is no real multilevel structure on such small networks. 
As soon as the instances get larger, we obtain a speedup of up to almost \emph{four orders of magnitude} (\Id{slashdot-zoo}). On average, \texttt{SCMLEvo} reaches the result of \texttt{GAEC+KLj} a factor 13.96 faster.  
Indeed, using parallelization helps to improve the edge cut of the final output, but also to achieve the result faster. Our parallel algorithm \texttt{SCMLEvoPar} computes the best result among all algorithms in all but two cases. 
It reaches a result equal to the result of \texttt{GAEC+KLj} faster by a factor~61.23 on average. 

At last we compare the performance of our algorithms against the recently proposed shared-memory parallel \texttt{ParCCML} algorithm. We run the algorithm using all cores of the machine. 
First, on 17 out of 20 instances, our sequential multi-level algorithm \texttt{SCML} computes a lower or equal edge cut (equal in two cases) than the \texttt{ParCCML} algorithm. The largest difference is observed on the \Id{cityscape} instances. Here, \texttt{SCML} computes cuts that are 15\%, 34\% and 30\% lower. On the instances on which \texttt{SCML} computes worse results, the differences are 1.5\%, 0.5\% and 0.4\%. Our sequential as well as our parallel evolutionary algorithm computes better results on all but one instance. In terms of running time, our sequential multilevel algorithm is 40\% faster than the \texttt{ParCCML} algorithm which is surprising as \texttt{ParCCML} runs on all 128 cores of our machine. This is in part due the size of the instances that are feasible for the evolutionary strategy and the overheads of parallel programming. With increasing size of the instances, there is an advantage in terms of running time of the parallel \texttt{ParCCML} algorithm. For example, on the \Id{cityscape} instances, the parallel \texttt{ParCCML} algorithm is a factor 3.2 faster than our sequential multi-level algorithm. Note that our evolutionary strategies use the multi-level algorithm as an algorithm to build the population. Therefore, while our (coarse-grained parallel) evolutionary algorithm is highly efficient, it can not surpass the performance of our multi-level strategy in terms of running time.

Lastly, we want to highlight that our memetic algorithm is more efficient in exploring the solution space than repeated executions of our multi-level strategy. This is evident from Figure~\ref{fig:convergenceplots} which shows how solution quality develops over time when \texttt{SCML} repetitions are run parallel on all cores of our machine as well as our parallel memetic algorithm \texttt{SCMLEvoPar}.
After the initial population has been created, one can clearly see that it is not sufficient to simply repeat the multilevel strategy for the same overall time as the memetic strategy to obtain similarly good solutions. 
\section{Conclusion}

In this study, we developed a novel approach to graph clustering, specifically tailored for signed graphs, which exhibit both positive and negative edge weights. We introduced a scalable multilevel algorithm that leverages label propagation and FM local search strategies to efficiently decompose a graph into clusters with minimal negative intra-cluster and maximal positive inter-cluster edge weights. Furthermore, we engineered a memetic algorithm that incorporates these strategies into a multilevel framework, enhancing the evolutionary process with robust recombination and mutation operations designed for signed graph clustering.
Our empirical results demonstrate that already sequentially our multilevel and memetic algorithms significantly outperform existing state-of-the-art algorithms in terms of both computational speed and the quality of the clustering results. An additional coarse-grained parallelization of our memetic algorithm further significantly reduces the time required to compute the clusterings.
Looking ahead, there are several avenues for further research. Enhancing the algorithm with more advanced local search mechanisms, such as those based on maximum flow models, could potentially yield even better clusterings. Additionally, extending our methodology to dynamically changing graphs and considering scalable shared-memory parallel algorithms based on deep multilevel schemes such as \cite{DBLP:conf/esa/GottesburenH00S21} could path the way for new applications.

\textbf{Acknowledgements.} We acknowledge support by DFG grant SCHU 2567/5-1. Moreover, we would like to acknowledge Dagstuhl Seminar 23331 on Recent Trends in Graph Decomposition.%
\bibliographystyle{abbrvnat}
\bibliography{compactfixed}

\vfill

\end{document}